\documentclass[
showpacs,
floatfix,
aps,
prl,
amsmath,
twocolumn,
superscriptaddress,
tightenlines
groupaddress,
]{revtex4}%{revtex4-1}
\usepackage{graphicx}
\usepackage{dcolumn}
\usepackage{amsmath}
\usepackage{amsfonts}
\usepackage{bm}
\usepackage{epsfig}
\usepackage{times}

\newcommand{\be}{\begin{equation}}
\newcommand{\ee}{\end{equation}}
\newcommand{\bea}{\begin{eqnarray}}
\newcommand{\eea}{\end{eqnarray}}

\newcommand{\wc}{\omega_{\rm c}}

\newcommand{\Rc}{R_c}

\newcommand{\vare}{\varepsilon}
\newcommand{\ve}{\varepsilon}

\def\gsm{\Gamma^{\rm (sm)}_{\rm ph}}
\def\gosc{\Gamma^{\rm (osc)}_{\rm ph}}

\def\eac{\epsilon_{\mathrm{ac}}}
\def\edc{\epsilon_{\mathrm{dc}}}
\def\eph{\epsilon_{\mathrm{ph}}}

\def\oc{\omega_{\mbox{\scriptsize {c}}}}

\def\rc{R_{\mbox{\scriptsize {c}}}}

\def\tauq{\tau_{\mbox{\scriptsize {q}}}}
\def\tautr{\tau_{\mbox{\scriptsize {tr}}}}

\newcommand{\req}[1]{Eq.~(\ref{#1})}
\newcommand{\reqs}[1]{Eqs.~(\ref{#1})}
\newcommand{\rref}[1]{(\ref{#1})}

\bibpunct{[}{]}{,\!}{n}{,}{,}
\begin{document}

\title{
Phonon-Induced Resistance Oscillations of Two-Dimensional Electron Systems 
Drifting with Supersonic Velocities
}

\author{I. A. Dmitriev}\thanks{Also at Ioffe Physical Technical Institute, 194021 St.~Petersburg, Russia}
  \affiliation{ Institut f\"ur Nanotechnologie, Karlsruher Institut f\"ur Technologie, 76021 Karlsruhe, Germany}
\author{ R. Gellmann}\thanks{Present address: Institute of Mechanics, University of Kassel, 34125 Kassel, Germany}
\affiliation{Inst. f\"ur Theorie der Kondensierten Materie, Karlsruher Institut f\"ur Technologie, 76128 Karlsruhe, Germany}
\author{M. G. Vavilov}
\affiliation{Department of Physics, University of Wisconsin, Madison, Wisconsin 53706, USA }
\date{\today}

\begin{abstract}
We present a theory of the phonon-assisted nonlinear dc transport of 2D electrons in 
high Landau levels. The nonlinear dissipative resistivity displays quantum magneto-oscillations governed by two parameters 
which are proportional to the Hall drift velocity $v_H$ of electrons in electric field and the speed of sound $s$.
In the subsonic regime, $v_H<s$, 
the theory quantitatively reproduces 
the oscillation pattern observed in recent experiments. 
We also find the $\pi/2$ phase change of oscillations across the sound barrier $v_H=s$.
In the supersonic regime, $v_H>s$, 
the amplitude of oscillations saturates with lowering temperature, while the subsonic region displays 
exponential suppression of the phonon-assisted oscillations with temperature.
\end{abstract}
\maketitle

\noindent A high-mobility two-dimensional electron gas (2DEG) in a weakly quantizing perpendicular magnetic field $B$ displays a rich variety of 
magneto-oscillation phenomena. These phenomena are defined  by the ratio of energy parameters and the Landau level spacing  $\hbar\oc$, with $\oc=eB/mc$, or by ratio of the spatial length scales and the cyclotron radius $\rc=v_F/\oc$, where $v_F$ ($p_F=mv_F$) is the Fermi velocity (momentum). 

Examples of magneto-oscillations include:
(\emph{i}) the microwave-induced resistance oscillations (MIRO)~\citep{miro:exp,miro:theory,ryzhii} and associated zero-resistance states (ZRS)~\citep{zrs:exp,zrs:theory}, controlled by the ratio of the  the microwave frequency $\omega$ and cyclotron frequency $\oc$,  $\eac\equiv\omega/\oc$;
(\emph{ii}) the Hall field-induced resistance oscillations (HIRO)~\citep{hiro:exp,VAG} and associated zero-differential resistance states (ZdRS)~\citep{zdrs} in the dc electric field $E$, governed by $\edc\equiv e E(2\rc)/\hbar \oc$;
(\emph{iii}) the phonon-induced resistance oscillations (PIRO)~\citep{piro:exp,zhang:2008,hatke:2009b,bykov:2010}, characterized by the ratio $\eph\equiv \omega_{\pi}/\oc$, where $\hbar \omega_{\pi}=2sp_F$ is
the energy of an acoustic  phonon  with momentum $2p_F$ and $s$ is the speed of sound. 
The above oscillations stem from commensurability of energy change in photon-- (MIRO); disorder-- (HIRO), and phonon--assisted (PIRO) scattering events and the $\hbar\wc$-periodic oscillations of the density of states (DoS).
MIRO and HIRO~\citep{miro:exp,miro:theory,zrs:exp,zrs:theory,hiro:exp,VAG,zdrs},
as well as their interplay~\citep{hatke:2008,khodas:2008,khodas:2010}, 
have been studied 
in detail over the past decade.

In this paper, we 
present a theory of a phonon-assisted transport in strong dc electric fields.  In the nonlinear dc-response, our results 
are consistent with the observed evolution of PIRO in the regime $\edc<\eph$ 
in recent experiment~\citep{zhang:2008}. We predict that  the phase change of oscillations across the sound barrier $\edc=\eph$, where the Hall drift velocity $v_H=cE/B$ equals to 
%%%
the speed of sound.  
We also analyze the temperature dependence of the phonon-assisted transport in different regimes.

\begin{figure}
\includegraphics[width=0.95\columnwidth]{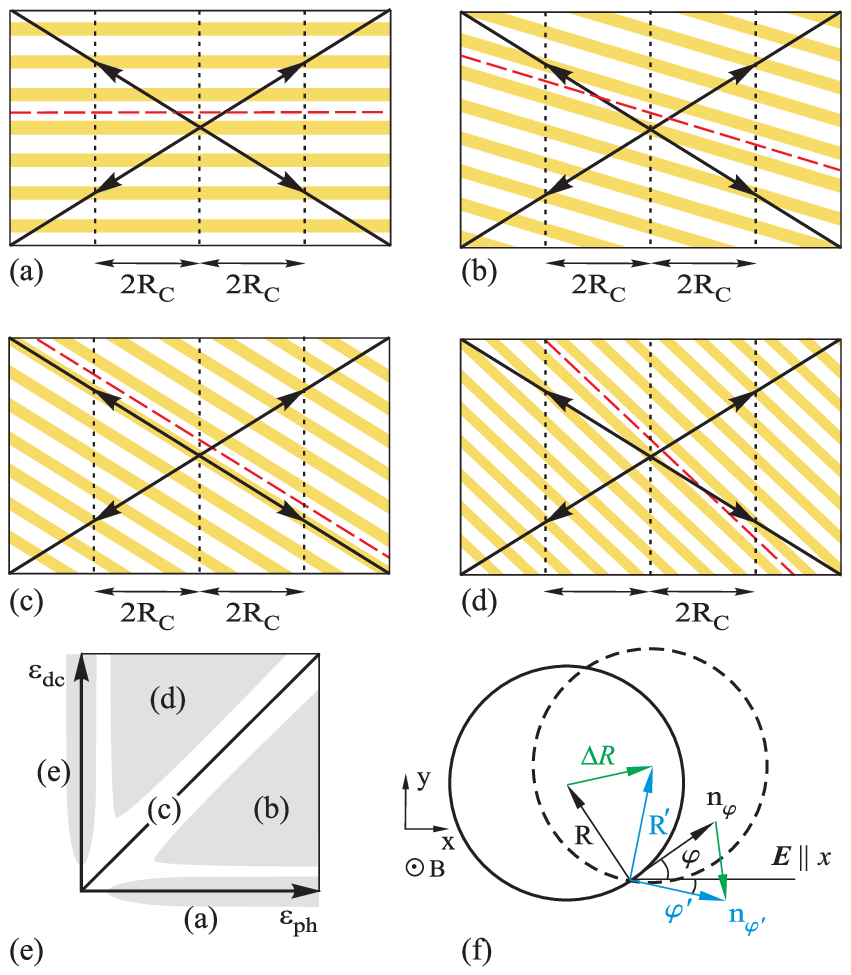}
\caption{Illustration of electron-phonon scattering processes in different regimes: (a) for $\edc=0$; (b) for $\edc<\eph$; (c) for $\edc=\eph$; (d) for $\edc>\eph$. Arrows show the change of the total energy of electron due to emission or absorption of a $2p_F$-phonon accompanied by the shift of the guiding center of the cyclotron orbit by $2R_C$ along or against the dc field. The arrows are directed along the sound cone (solid lines), while the strength of dc field determines the tilt of Landau levels (maxima of oscillatory DoS \req{nu} are marked by yellow stripes). Dashed line marks the position of the Fermi level. The position of four regions Fig.~\ref{fig:phiro-regions}a-d in the $\eph-\edc$ plane is shown in Fig.~\ref{fig:phiro-regions}e. Fig.~\ref{fig:phiro-regions}f illustrates 
the real-space shift $\Delta R$ of the cyclotron guiding center 
and the change in the electron momentum for scattering from 
$\varphi$ to $\varphi'$.
 }
 \label{fig:phiro-regions}
\end{figure}

{\it Model.} For a homogeneous 2DEG in a 
region of 
classically strong $B$, when $\wc$ exceeds the transport scattering rate $1/\tautr$,
the dissipative current $j_x$ in response to an %homogeneous 
electric field $E$ (pointing along $x$ direction, $\varphi=0$) is %given by the expression
\begin{equation}
\label{current-gen}
j_x E = 2 \nu_0\int \!  W_{\varphi,\varphi'} P_{\varphi,\varphi'} (\ve)d\ve \frac{d\varphi d\varphi'}{4\pi^2} \,.
\end{equation} 
Here $\nu_0=m/2\pi$ is the DoS at zero magnetic field (hereafter we put $\hbar=1$) and $P_{\varphi,\varphi'}(\ve)$ denotes the semiclassical scattering 
rate of 
an electron with energy $\ve$ with  the change in its momentum direction from $\varphi$  to $\varphi'$:
\begin{equation}
\begin{split}
&P_{\varphi,\varphi'}(\ve)=\tau_{\varphi-\varphi'}^{-1} \cal{M} (\ve,\ve+W_{\varphi,\varphi'}) \\
& + \frac{g^2|\omega_{\varphi-\varphi'} |}{2}  
N_{\omega_{\varphi-\varphi'}}\,
{\cal M}(\ve,\ve+W_{\varphi,\varphi'}+|\omega_{\varphi-\varphi'}|)\\
&+ \frac{g^2|\omega_{\varphi-\varphi'}|}{2}  
(N_{\omega_{\varphi-\varphi'}}+1)\,
{\cal M}(\ve,\ve+W_{\varphi,\varphi'}-|\omega_{\varphi-\varphi'}|).
\end{split}
\label{P}
\end{equation}
In scattering of static disorder, represented by the first line in \req{P}, electron 
energy $\ve$ changes by the electrostatic work $W_{\varphi,\varphi'}=eE\Rc(\sin\varphi-\sin\varphi')$
for displacement $\Delta \bm{R}$ of electron cyclotron trajectory in electric field, see Fig.~\ref{fig:phiro-regions}f. 
The second (third) line in \req{P} describes processes with change in electron electrostatic energy 
$W_{\varphi,\varphi'}$ and absorption (emission) of a
phonon with energy $|\omega_{\varphi-\varphi'}|$, where 
$\omega_{\varphi-\varphi'}=2p_F s \sin(\varphi-\varphi')/2$ for $s\ll v_F$.
The Planck's function $N_{\omega}=1/[\exp(|\omega|/T)-1]$ accounts for the thermal occupation of phonon modes,
and $g$ is the dimensionless electron-phonon coupling constant~\citep{note}.

Factors ${\cal M}(\ve,\ve')=\tilde\nu(\ve)\tilde\nu(\ve')f(\ve)[1-f(\ve')]$ in \req{P} contain the Fermi function $f(\vare)=1/\{\exp[(\vare-\mu)/T]+1\}$ and the product of 
the DoS of initial and final %scattering 
states. The DoS in the experimentally relevant region of weak magnetic fields, $\tauq\wc\lesssim 1$, 
has the form 
\be
\label{nu}
\tilde\nu(\ve)\equiv\nu(\ve)/\nu_0=1-2\lambda\cos(2\pi\ve/\wc),
\ee
where $\lambda=\exp(-\pi/\wc\tau_{\rm q})$ and $\tauq$ is the quantum scattering time off disorder.

Using Eqs.~(\ref{current-gen})-(\ref{nu}) and assuming $T \tauq\gtrsim 1$, we obtain
\begin{equation}
\label{current-gamma}
j_x E = \frac{\rho_D E^2\tautr}{\rho_H^2}  \Big(\Gamma_{\rm dis} +
\Gamma_{\rm ph}^{\rm (sm)} +\Gamma_{\rm ph}^{\rm (osc)} \Big),
\end{equation}
where $\rho_D=\pi\nu_0/(e^2N_e\tautr)$ and $\rho_H=B/(ecN_e)$ are the Drude and Hall resistivities, $N_e=mv_F^2\nu_0$ is the electron surface density. The effective scattering rate of disorder was obtained in \citep{VAG}: $\Gamma_{\rm dis}(\edc)=\tautr^{-1}-2\lambda^2 F''(\pi\edc)$, with $F(x)=\sum J_n^2(x)/\tau_n$
and $J_n(x)$ standing for the Bessel functions; the scattering rate $\tau^{-1}_{\varphi-\varphi'}=\sum\tau^{-1}_n e^{i n (\varphi-\varphi')}$ is presented by its angular harmonics $1/\tau_n$ ($\tau_0\equiv\tauq$). 

Here we focus on the phonon-assisted scattering rates
\bea
\nonumber
&&\left\{
\begin{array}{c}
\Gamma_{\rm ph}^{\rm (sm)}%(\edc,\eph)
\\  \Gamma_{\rm ph}^{\rm (osc)}%(\edc,\eph)
\end{array}\right\} = g^2T\int \left\{
\begin{array}{c}
1\\  2\lambda^2\cos\displaystyle\frac{W_{\varphi,\varphi'}-\omega_{\varphi-\varphi'}}{\wc/2\pi}
\end{array}\right\}
\\\label{rates}
&&\times
(\sin\varphi-\sin\varphi')^2\Lambda\Big(\displaystyle\frac{\omega_{\varphi-\varphi'}}{2T}, \frac{W_{\varphi,\varphi'}}{2T}\Big)\frac{d\varphi d\varphi'}{4\pi^2}
\, ,
\eea
where $\gsm(\edc,\eph)$ and $\gosc(\edc,\eph)$ denote the smooth and oscillatory components of the electron-phonon scattering rate, and $\Lambda(x,y)=S(x)S(x-y)/S(y)$ is written in terms of $S(x)\equiv{x}/{\sinh x}$. We first study the behavior of $\Gamma^{\rm (sm)}_{\rm ph}$ and $\Gamma^{\rm (osc)}_{\rm ph}$ in the experimentally relevant  %high temperature 
limit $T\gg \wc$ and then analyze how these rates evolve with lowering $T$.

\begin{figure}
\includegraphics[width=0.95\columnwidth]{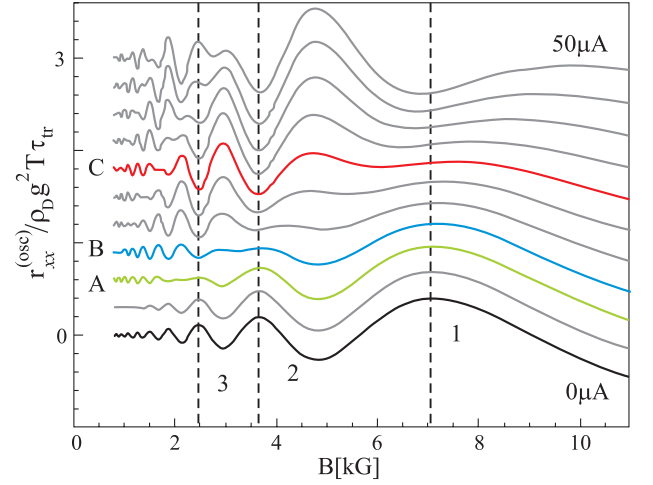}
\caption{$r_{xx}^{(\rm osc)}$ vs $B$ for several values of current $I$ varying from 0 $\mu$A to 50 $\mu$A in 5 $\mu$A steps. Traces are vertically offset for clarity.
Traces (A) [10$\mu$A], (B) [15$\mu$A], and (C) [30$\mu$A] correspond approximately to the values of $I$ at which 3rd, 2nd, and 1st PIRO peaks disappear before evolving into minima at higher currents. Very similar values (A) [10$\mu$A], (B) [14$\mu$A], and (C) [28$\mu$A] were found in the experiment (see Fig.~2 of Ref.~ \cite{zhang:2008}).}
 \label{fig:2}
\end{figure}

\emph{Main results.}
At high $T\gtrsim W,\omega$, we approximate $\Lambda(\omega/2T,W/2T)\approx 1$ in \req{rates} and %immediately
obtain constant $\gsm = g^2T$ for the smooth part of electron-phonon transport scattering rate. The oscillatory part, which represents the effect of the Landau quantization, reduces to
\begin{equation}
\begin{split}
\gosc & = 8\lambda^2 g^2T\int_0^{2\pi}\frac{d\varphi_+d\varphi_-}{4\pi^2}\sin^2\varphi_-\cos^2\varphi_+\\
&\times \cos\left[2\pi\sin\varphi_-(\eph-\edc\cos\varphi_+)
\right].
\end{split}
\label{gosc-gen}
\end{equation}

The behavior of $\Gamma^{\rm (osc)}_{\rm ph}(\edc,\eph)$ is different in 
parametric regimes, illustrated in  Fig.~\ref{fig:phiro-regions}e.
%Case A is 
In the limit of weak electric fields, $\edc\ll 1$, the current is linear in the applied dc field and
\be\label{a}
\gosc(0,\eph)=2\lambda^2 g^2T[J_0(2\pi\eph)-J_2(2\pi\eph)].
\ee
The oscillations with $\eph=\omega_\pi/\wc$ in Eq.~\eqref{a} are driven by commensurability between the phonon energy $\omega_\pi=2p_Fs$ and the period of DoS oscillations $ \wc$, see Fig.~\ref{fig:phiro-regions}a. For $\eph\gtrsim 1$, Eq.~\eqref{a} gives $\gosc(0,\eph)=(4\lambda^2 g^2T/\pi\sqrt{\eph}) \cos(2\pi\eph-\pi/4)$. %We note that t
This regime was studied previously by Raichev \cite{raichev:2009}. The difference in the oscillations phase of our result and that of Ref.~\cite{raichev:2009} is due to a
different choice of the photon scattering form-factor in narrow vs. wide quantum wells.

A strong Hall field tilts the Landau levels and changes the commensurability conditions for electron-phonon scattering as shown in Fig.~\ref{fig:phiro-regions}b-d, resulting in oscillations with combined parameters $\epsilon_\pm\equiv\edc\pm\eph$. Equivalently, this effect can be viewed as a result of the Doppler shift of the phonon modes in the frame moving with the Hall velocity $v_H$
in $y$ direction. In the moving frame, the electric field is absent while the phonon speed
changes as $s\to s-v_H\cos\varphi_+$ in the argument of cosine in the second line of Eq.~\eqref{gosc-gen}. The latter formulation is particularly useful in the case of complicated spectrum of phonons.
%, considered in Ref.~\cite{raichev:2009}. 

At $\edc\gg 1$, the rate $\gosc$ can be approximated as
\begin{eqnarray}\label{eq:goscSF}
\gosc(\edc,\eph)&=&-\frac{2}{\pi^2}\lambda^2 g^2T\frac{\partial^2 \Phi(\edc,\eph)}{\partial \edc^2},\\
\nonumber
\Phi(\edc,\eph)&=&\frac{1}{\sqrt{\edc}}\sum_\pm B_{\rm sign \epsilon_\pm}(|\epsilon_\pm|);\\
\nonumber
\quad B_{\pm1}(z)&=&\sqrt{{z}/{8}}[J^2_{-1/4}(\pi z)\pm J^2_{1/4}(\pi z)].
\end{eqnarray}
This expression can be simplified in three cases when electron drifts with subsonic velocity ($v_H<s$ or $\edc<\eph$, Fig.~\ref{fig:phiro-regions}b);  velocity of sound ($v_H=s$ or $\edc=\eph$, Fig.~\ref{fig:phiro-regions}c); and supersonic velocity ($v_H>s$ or $\edc> \eph$, Fig.~\ref{fig:phiro-regions}d):%. The corresponding asymptotic expansions of \req{eq:goscSF} for $\edc\gg 1$ are
\begin{equation}\begin{split}
& \gosc(\edc,\eph)=4\lambda^2g^2T\\
&\times \left\{
  \begin{array}{ll}\displaystyle 
    \left(\frac{\sin 2\pi\epsilon_+}{\pi^2\sqrt{\edc\epsilon_+}}+\frac{\cos 2\pi\epsilon_-}{\pi^2\sqrt{\edc|\epsilon_-|}}\right), & \hbox{$\edc\ll \eph$;} \\
    \displaystyle 
    \left(\frac{1}{3\sqrt{\pi\edc}\Gamma^2(3/4)}+\frac{\sqrt{2}}{2\pi^2\edc}\right), & \hbox{$\edc=\eph$;} \\
   \displaystyle 
    \left(\frac{\sin 2\pi\epsilon_+}{\pi^2\sqrt{\edc\epsilon_+}}+\frac{\sin 2\pi\epsilon_-}{\pi^2\sqrt{\edc\epsilon_-}}\right), & \hbox{$\edc\gg \eph$.}
  \end{array}
\right.
\end{split}
\label{3}
\end{equation}

Evolution of PIRO in subsonic sector $\edc<\eph$ was studied experimentally in Ref.~\cite{zhang:2008}, where
 the differential resistivity $r_{xx}=\partial E/\partial j_x\simeq \rho_H^2 \partial j_x/\partial E$ 
was measured. The PIRO contribution to the differential resistivity $r_{xx}^{(\rm osc)}$ can be evaluated using \reqs{current-gamma} and \rref{rates} as $r_{xx}^{(\rm osc)}/\rho_D=\partial/\partial\edc[\edc \Gamma_{\rm ph}^{(\rm osc)}(\edc,\eph)]$.

In Fig.~\ref{fig:2} we plot $r_{xx}^{(\rm osc)}$, calculated according to Eqs.~\eqref{nu} and \eqref{gosc-gen} with $\tauq=8$~ps as a function of $B=B_0/\eph$ for several values of the total current  $I=j_x w=I_0\edc/\eph$. Here $w=50~\mu$m is the Hall bar width, $B_0=2p_F s m c/e=7.67$~kG and $I_0=e N_e s w=223~\mu$A are the reference field and current; values of the parameters $s$, $N_e$, and $w$ were taken from Ref.~\cite{zhang:2008}. This theoretical result quantitatively reproduces the experimentally observed evolution of PIRO  with increasing $I$ \cite{zhang:2008} without fitting parameters.

The important feature of Eq.~\eqref{3} is the change in the phase of oscillations across the sound barrier $v_H=s$ (see Fig.~\ref{fig:phiro-regions}c), which calls for an experimental verification.
In the supersonic sector $\edc>\eph$, see Fig.~\ref{fig:phiro-regions}d,  the scattering along electric field always results in gain in the Landau level index,% (kinetic energy), 
while scattering against electric field reduces the Landau level index.

In the extreme  limit of strong $B$ or soft phonons $\eph\ll 1$ (see region (e) in Fig.~\ref{fig:phiro-regions}e), Eq.~\eqref{gosc-gen} gives
\begin{equation}
\frac{\gosc(\edc,0)}{2\lambda^2 g^2 T}\approx J_0^2(\pi\edc)-J_1^2(\pi\edc)-J_0(\pi\edc)J_2(\pi\edc).
\label{e}
\end{equation}
Here,  the electron-phonon scattering behaves as a scattering off nearly static impurities and
the functional form Eq.~(\ref{e}) 
coincides with that of HIRO, presented above as $F''(\pi\edc)$, for the 
specific case of isotropic disorder scattering, $\tau_\theta\equiv \tau_0$.

\begin{figure}
\includegraphics[width=0.95\columnwidth]{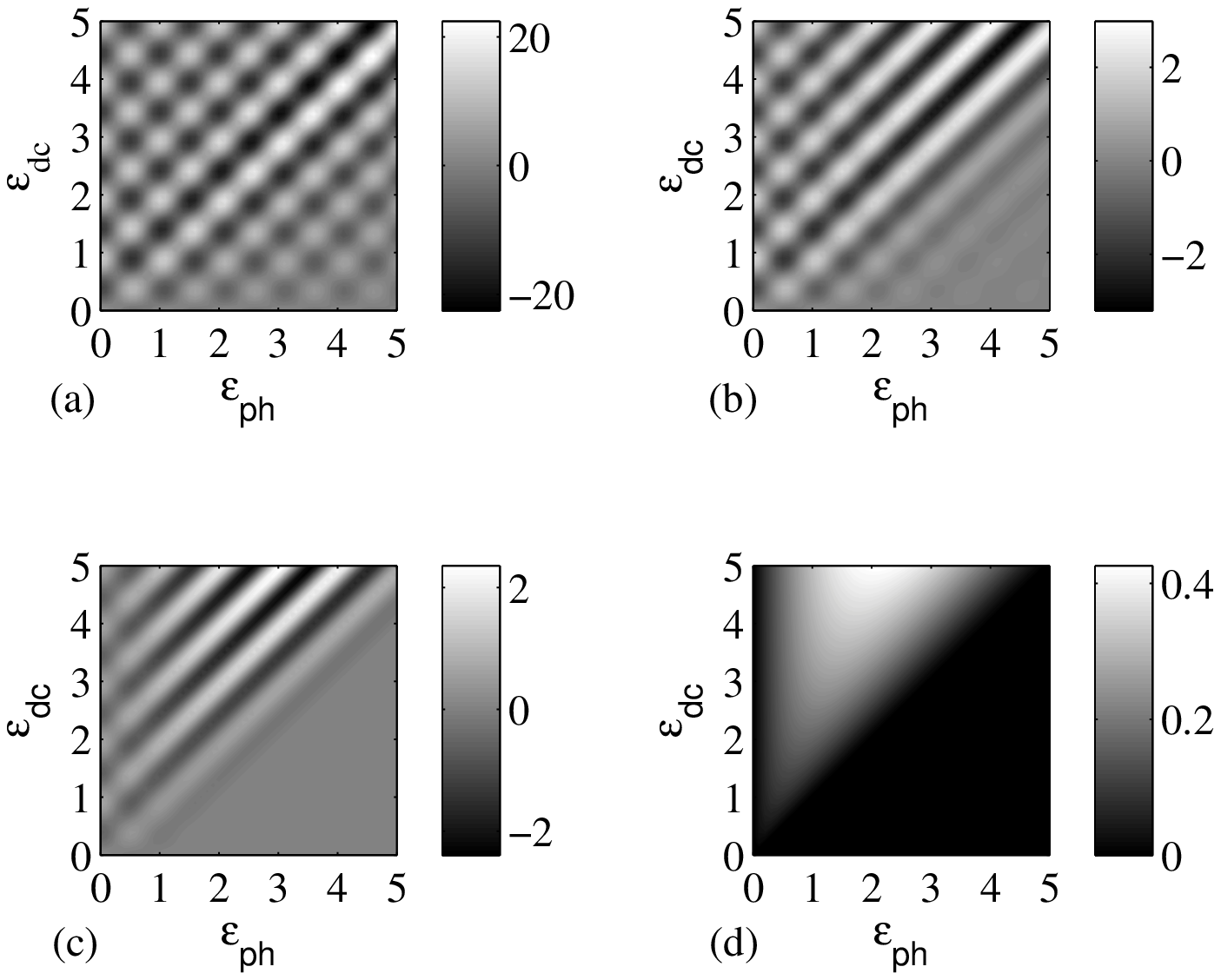}
\caption{(a)-(c): Oscillatory part of 
the phonon-assisted differential resistivity $r^{(\rm osc)}_{xx}$ in units $  2\lambda^2 g^2 \rho_D\wc\tautr $ 
calculated using \reqs{current-gamma}, \rref{rates} for the case of high [$T=5\wc$, panel (a)], intermediate
[$T=0.7 \wc$, panel (b)], and low [$T=0.25 \wc$, panel (c)]  temperature. Panel (d): Smooth part of phonon-assisted scattering rate $\gsm(\edc,\eph)$ at zero temperature, measured in units of $g^2\wc$, \req{gsmzeroT}.
}
 \label{fig:3}
\end{figure}

As temperature $T$ decreases,  the phonon-assisted magneto-oscillations in differential resistivity become suppressed completely  in the subsonic regime.  
However, in the supersonic regime only $\epsilon_+$ oscillations vanish, while 
amplitude of $\epsilon_-$ oscillations  saturates, see Figs.~\ref{fig:3}a-c.

This behavior can also be inferred from the analytical results for $\gosc$, \req{rates},
obtained for low temperatures $T\lesssim 2p_F s$, but still $T\gtrsim  \wc$. 
%In the subsonic regime, 
For $\edc<\eph$, we have
\begin{equation}\begin{split}
& \gosc(\edc,\eph)  =\frac{4\lambda^2g^2 \wc\eph}{\pi^2\edc^{3/2}}\exp(-2|\epsilon_-|\wc/T)\\
&\times \left[\sqrt{|\epsilon_-|}\cos2\pi\epsilon_-+\sqrt{\epsilon_+} e^{-2\edc\wc/T}\sin 2\pi\epsilon_+\right].
\end{split}
\end{equation} 
$\gosc$ is exponentially small for $\epsilon_-<0$, but as the sound barrier is reached, a non-exponential contribution emerges
\begin{equation}\begin{split}
& \gosc(\edc,\eph)  =2\lambda^2g^2 \wc\\
&\times \left[\frac{T/\wc}{3\sqrt{\pi\edc}\Gamma^2(3/4)}+\frac{\sqrt{8}}{\pi^2} e^{-2\edc\wc/T} \sin 4\pi\edc \right].
\end{split}
\end{equation} 
In the supersonic regime, $\epsilon_-=\edc-\eph>0$, we obtain
\begin{equation}
\begin{split}
& \gosc(\edc,\eph)  =\frac{2\lambda^2g^2 \wc\eph}{\pi^2\edc^{3/2}}\\
&\times \left[\sqrt{\epsilon_-}\sin 2\pi\epsilon_-+\sqrt{\epsilon_+} e^{-2\eph\wc/T}\sin 2\pi\epsilon_+\right].
\end{split}
\end{equation}

A similar behavior also occurs with the smooth  component of the electron-phonon  rate $\gsm$: it vanishes in the subsonic regime and saturates in the supersonic regime, see Fig.~\ref{fig:3}d. For $T=0$, the smooth component
%we obtain 
%mv: the following expression for the smooth part of the electron-phonon transport rate at $T\to 0$
\begin{equation}
\gsm(\edc,\eph)=\frac{8g^2\wc\eph}{3\pi^2}
\left[\arccos\frac{\eph}{\edc}-\frac{\eph}{\edc^2}\sqrt{\epsilon_+\epsilon_-}\right]
\label{gsmzeroT}
\end{equation}
for $\edc>\eph$ and $\gsm(\edc,\eph)=0$ otherwise. 
%mv: This contribution vanishes in the subsonic regime and remains finite in the supersonic regime, when $\edc>\eph$. 
At finite but low temperature, $\gsm(\edc,\eph)$ remains small in the subsonic sector and grows continuously across the sound barrier.

Such contrast in low-temperature behavior of electron-phonon scattering rates $\gsm$ and $\gosc$ in the subsonic and supersonic regimes can be understood from the structure of \req{P}. 
At low temperatures $T\lesssim 2p_F s$,
the occupation number of phonon modes with energy $\omega\sim 2 p_F s$ is exponentially small.  Therefore, the electron scattering off thermal phonons becomes negligible. At the same time, the spontaneous emission of phonons depends only on the  combination $f(\vare_{in})[1-f(\vare_f)]$ of electron occupation numbers $f(\vare)$ at initial ($\vare_{in}$) and final ($\vare_f$) energies. For emission processes, this combination is also exponentially small in equilibrium, but may remain finite in a system subject to electric fields. When the Hall drift exceeds the speed of sound, $\vare_f$ for electron scattering with increase in electrostatic energy is larger than $\vare_i$ even for the process with phonon emissions, see Fig.~\ref{fig:phiro-regions}d, and the spontaneous phonon emission takes place.

\emph{Conclusions.}
We developed a theory of phonon-assisted magneto-oscillations in strong dc electric fields.  In the limit of a weak Hall field, $\edc\ll 1 \ll\eph$, or in the limit of quasi-elastic phonon scattering, $\eph\ll 1 \ll\edc$, the theory  is consistent with the known results for PIRO~\citep{raichev:2009} and HIRO~\citep{VAG}. In the nonlinear dc-response $\edc<\eph$, the theory  reproduces the evolution of PIRO observed in recent experiment, cf. Ref.~\citep{zhang:2008} and Fig.~\ref{fig:2}. We find that the phase of oscillations changes across the sound barrier at $\edc=\eph$,
where the Hall drift velocity $v_H=cE/B$ reaches 
the speed of sound  $s$, which also warrants an experimental verification.
In the supersonic regime $\edc>\eph$, the amplitude of oscillations saturates at finite value for 
very low temperatures $T\ll \omega_\pi$, while in the subsonic sector $\edc<\eph$ the phonon-assisted current decays exponentially with lowering $T$, as the number of thermal phonons decreases. At zero temperature, the smooth part of the phonon-assisted nonlinear conductivity also exhibits saturation to a finite value in the supersonic regime, while vanishes in the subsonic regime.

We gratefully acknowledge D.~N.~Aristov, A.~A.~Bykov, M.~Khodas, A.~J.~Millis, A.~D.~Mirlin, D.~G.~Polyakov, I.~V.~Protopopov, S.~A.~Vitkalov, and M.~A.~Zudov for discussions.
ID was supported by the DFG, by the DFG-CFN, by Rosnauka Grant no. 02.740.11.5072, by the RFBR, and MV was supported by NSF Grant No. DMR-0955500.

% % %\bibliography{bibPHIRO.bib}

\begin{thebibliography}{17}
\bibitem{note}{We consider interaction with 2D isotropic deformational phonons
  justified if $z$-component of relevant phonon momentum is negligible (i.e.
  the width of the quantum well where 2DEG is confined $b\gg k_F^{-1}, s/T$).
  In this case, $g^2=m D^2/\rho b s^2$ in terms of the mass density $\rho $ and
  deformation-potential $D$ of the host crystal. General case will be presented
  elsewhere.}
\end{thebibliography}

\end{document}